\documentclass[12pt]{article}

\usepackage{scicite}
\usepackage{graphicx}

\usepackage{times}

\newenvironment{sciabstract}{%
\begin{quote} \bf}
{\end{quote}}

\newcounter{lastnote}
\newenvironment{scilastnote}{%
\setcounter{lastnote}{\value{enumiv}}%
\addtocounter{lastnote}{+1}%
\begin{list}%
{\arabic{lastnote}.}
{\setlength{\leftmargin}{.22in}}
{\setlength{\labelsep}{.5em}}}
{\end{list}}

\title{Itinerant Ferromagnetism in a Fermi Gas of Ultracold Atoms}

\author
{ Gyu-Boong Jo$^{1\ast}$, Ye-Ryoung Lee$^{1}$, Jae-Hoon Choi$^{1}$, Caleb A. Christensen$^{1}$, \\
 Tony H. Kim$^{1}$, Joseph H. Thywissen$^{2}$, David E. Pritchard$^{1}$, and Wolfgang Ketterle$^{1}$\\
\\
\normalsize{$^{1}$MIT-Harvard Center for Ultracold Atoms, Research
Laboratory of Electronics,}\\
\normalsize{ Department of Physics, Massachusetts
Institute of Technology, Cambridge, MA 02139, USA}\\
\normalsize{$^{2}$Department of Physics, University of Toronto, Toronto, Ontario M5S1A7, Canada}\\
\\
\normalsize{$^\ast$To whom correspondence should be addressed; E-mail: gyuboong@mit.edu}
}

\date{}

\begin{document}


\baselineskip24pt


\maketitle


\begin{sciabstract}

Can a gas of spin-up and spin-down fermions become ferromagnetic due to repulsive interactions?
This question which has not yet found a definitive theoretical answer was addressed in an experiment
 with an ultracold two-component Fermi gas.  The observation of non-monotonic behavior of lifetime,
 kinetic energy, and size for increasing repulsive interactions provides strong evidence for a phase
 transition to a ferromagnetic state.  It implies that itinerant ferromagnetism of delocalized fermions
 is possible without lattice and band structure and validates the most basic model for ferromagnetism introduced by Stoner.

\end{sciabstract}



Magnetism is a macroscopic phenomenon with its origin deeply rooted in quantum
mechanics.  In condensed matter physics, there are two paradigms
for magnetism:  localized spins interacting via tunnelling, and
delocalized spins interacting via an exchange energy. The latter
gives rise to itinerant ferromagnetism which is responsible for
the properties of transition metals like cobalt, iron and nickel.  Both
kinds of magnetism involve strong correlations and/or strong
interactions and are not yet completely understood.  For localized spins, major open questions
include the interplay of magnetism with d-wave superfluidity and
frustrated spin materials.  For itinerant ferromagnetism~\cite{FBloch29,stoner33,FM1d,PhysRevB.60.R13977,Vollhardt,PhysRevLett.75.4678,tanaka:116402}, phase
transition theories are still qualitative.

We implement the Stoner model, a textbook Hamiltonian for itinerant ferromagnetism~\cite{snoke}, using a two-component gas of free fermions with
short-range repulsive interactions, which can capture the essence of the screened Coulomb interaction in
electron gases~\cite{snoke}.  However, there is no proof
so far that this simple model for ferromagnetism is consistent
when the strong interactions are treated beyond mean-field
approaches.  It is known that this model fails in one
dimension where the ground state is singlet for arbitrary
interactions, or for two particles in any dimension~\cite{FM1d}.
Here, cold atoms are used to perform a quantum simulation of this
model Hamiltonian in 3D and show experimentally that it leads to
a ferromagnetic phase transition~\cite{stoner33}. This model is also realized in helium-3~\cite{Vollhardt84}, but it turns
into solid and not into a ferromagnetic phase at high pressure. It has also been applied
to neutrons in neutron stars~\cite{neutronstar}.

So far, magnetism in ultracold gases has been studied only for spinor~\cite{Stenger98,KurnFM}
 and dipolar~\cite{dipolarbec} Bose-Einstein condensates.
In these cases, magnetism is driven by weak spin-dependent interactions
which nevertheless determine the structure of the condensate due to
a bosonic enhancement factor. In contrast, here we simulate quantum magnetism in a strongly interacting
Fermi gas.

An important recent development in cold atom science has been the
realization of superfluidity and the BEC-BCS crossover in strongly
interacting two-component Fermi gases near a Feshbach resonance~\cite{verennaKetterle}.
  These phenomena occur for attractive
interactions for negative scattering length and for bound
molecules (corresponding to a positive scattering length for two
unpaired atoms).  Very little attention has been given to
the region for atoms with strongly repulsive interactions.  One reason is
that this region is an excited branch, which is unstable against near-resonant three-body
recombination into weakly-bound molecules.  Nevertheless,  many
theoretical papers have proposed a two-component Fermi gas near a
Feshbach resonance as a model system for itinerant ferromagnetism
~\cite{PhysRevA.56.4864,Reatto00,Tosi00,PhysRevA.66.043611,PhysRevLett.95.230403,Pcoleman09,WuFM08,thywissen09}
assuming that the decay into molecules can be sufficiently  suppressed.  Another open question which has not been
addressed is the possibility of a fundamental limit for repulsive interactions.
 Such a limit due to unitarity or many-body physics may be lower than the value required for
  the transition to a ferromagnetic state. We show that this is not the case, and that
there is a window of metastability where the onset of ferromagnetism can be observed.

A simple mean-field model captures many qualitative features of
the expected phase transition, but is not adequate for a
quantitative description of the strongly interacting regime. The
total energy of a two-component Fermi gas of average density $n$
(per spin component) in a volume $V$ is given by
$E_{F}2Vn[\frac{3}{10}\{(1+\eta)^{5/3}+(1-\eta)^{5/3}
\}+\frac{2}{3\pi}k_{F}a(1+\eta)(1-\eta)]$ where $E_{F}$ is the
Fermi energy of a gas, $k_F$ the Fermi wavevector of a gas, $a$ the scattering
length characterizing short-range interactions between the two
components, and $\eta=\Delta{n}/n=(n_1-n_2)/(n_1+n_2)$ magnetization of the Fermi gas.
The local magnetization of the Fermi gas is non-zero when the gas
separates into two volumes, where the densities $n_1$ and $n_2$ of
the two spin states differ by $2 \Delta{n}$.
Note that we study an ensemble in which the number of atoms in each spin state is conserved. This is equivalent to a free electron gas at zero external magnetic field where the total
magnetization is zero. The interaction term
represents any short-range spin-independent potential.
 When the gas is fully polarized, it avoids the repulsive interaction, but increases its
kinetic energy by a factor of $2^{2/3}$.  The phase transition
occurs when the minimum in energy is at non-zero magnetization
(Fig. 1A) at $k_{F}a={\pi}/2$. This onset was discussed in the
context of phase-separation in a two-component Fermi gas~\cite{PhysRevA.56.4864,Reatto00,Tosi00,PhysRevA.66.043611}.
  Fig. 1B shows several consequences of the phase transition for a system at constant
pressure.  First, for increasing repulsive interactions, the gas
expands, lowering its density and Fermi energy; kinetic energy is therefore reduced. When the gas enters the ferromagnetic phase,
kinetic energy increases rapidly due to the larger local density
per spin state. Furthermore, the volume has a maximum value at the phase transition.
This can be understood by noting that pressure in our model is $(2/3) E_{kin}/V + E_{int}/V$, where
$E_{kin}$ is kinetic energy and $E_{int}$ interaction energy. At the
phase transition, the system increases its kinetic energy and reduces its interaction
energy, thus reducing the pressure. This maximum in pressure at constant volume turns
into a maximum in volume for a system held at constant pressure, or in a trapping potential.
We have observed three predictions of this model: the onset of local magnetization
through the suppression in inelastic collisions, the minimum in kinetic energy,
and the maximum in the size of the cloud. These qualitative features are generic for the ferromagnetic
phase transition and should be present also in more advanced models~\cite{PhysRevLett.95.230403}.

We start with an atom cloud
consisting of an equal mixture of $^{6}$Li atoms in the lowest two
hyperfine states, held at 590 G in an optical dipole trap with
additional magnetic confinement~\cite{exp}. The number of atoms per spin state $\sim6.5\times10^{5}$
corresponds to a Fermi temperature $T_F$ of $\sim$1.4~$\mu$K. The effective temperature $T$ could
be varied between  $T/T_{F}=0.1$ and $T/T_{F}=0.6$ and was determined right after the field ramp by fitting the spatial
distribution of the cloud with a finite temperature
Thomas-Fermi profile. Note that $k_{F}^{\circ}$ is the Fermi wavevector of the non-interacting
gas calculated at the trap center. Applying the procedure discussed in
Ref.~\cite{Thomas05} to repulsive interactions, we estimate that
the real temperature is $\sim$20\% larger than the effective one.
The effective temperature did not depend on $k_{F}^{\circ}a$ for $k_{F}^{\circ}a < 6$.
At higher temperatures, additional shot-to-shot noise was caused by large fluctuations in the
atom number. From the starting point at 590~G,
the magnetic field was increased towards the Feshbach resonance at
834 G, thus providing adjustable repulsive interactions. Due to the limited lifetime of
the strongly interacting gas, it was necessary to apply the fastest possible field ramp, limited
to 4.5~ms by eddy currents. The ramp time is approximately equal to the inverse of the axial trap frequency~\cite{exp}
 and therefore only marginally adiabatic.  Depending
on the magnetic field during observation, either atoms or atoms
and molecules were detected by absorption imaging as described in Fig.~S1~\cite{PhysRevLett.91.250401}.

The emergence of local spin polarization can be observed by the
suppression of (either elastic or inelastic) collisions, as the
Pauli exclusion principle forbids collisions in a fully polarized
cloud.  We monitor inelastic three-body collisions which convert
atoms into molecules.  The rate (per atom) is proportional to $f(a,T)
n_1 n_2$ or $f(a,T) n^2(1-\eta^2)$ and is
therefore a measure of the magnetization $\eta$. For $k_{F}a \ll 1$, the rate coefficient
$f(a,T)$ is proportional to $a^6 \max(T,T_F)$~\cite{PhysRevLett.94.213201}.
This rate can be observed by monitoring the initial drop in the number of
atoms during the first 2~ms after the field ramp.  We
avoided longer observation times since the increasing molecule
fraction could modify the properties of the sample.

Fig.~2 shows a sharp peak in the atom loss rate around $k_{F}^{\circ}a \simeq
2.5$ at $T/T_{F} = 0.12$ indicating a transition in the sample to a state with
local magnetization.  The gradual decrease is consistent with the
inhomogeneous density of the cloud where the transition occurs
first in the center.  The large suppression of the loss rate
indicates a large local magnetization of the cloud.

The kinetic energy of the cloud was determined by suddenly
switching off the optical trap and the Feshbach fields right after the field ramp and then
imaging state $|1\rangle$  atoms at zero field  using the cycling
transition after a ballistic expansion time of
${\bigtriangleup}_{tof}=4.6$~ms.  The kinetic energy was obtained
from the Gaussian radial width $\sigma_x$ as
$E_{kin}=\frac{3m{\sigma_x}^2}{2{{\bigtriangleup}_{tof}}^2}$ where
$m$ is the mass of the $^{6}$Li atom.  Fig.~3 demonstrates a minimum of
the kinetic energy at $k_{F}^{\circ}a \simeq 2.2$ for the coldest temperature, $T/T_{F} = 0.12$, nearly coinciding with the onset
of local polarization. The peak in the atom loss rate occurs slightly later than the minimum of kinetic energy, probably
because $f(a,T)$ increases with $a$~\cite{thywissen09}.
Since the temperature did not change around $k_{F}^{\circ}a \simeq 2.2$, the
increase in kinetic energy is not caused by heating, but by a
sudden change in the properties of the gas, consistent with the
onset of ferromagnetism.  The observed increase in kinetic energy
is $\sim$20 \% at $T/T_{F} = 0.12$, smaller than the value $(2^{2/3}-1)=0.59$ predicted
for a fully polarized gas. This discrepancy could be due to
the absence of polarization or partial polarization in the wings of the cloud.
Also, it is possible that the measured kinetic energy of the strongly interacting gas before
the phase transition includes some interaction energy if the Feshbach fields are not suddenly switched off.
For the current switch-off time of $\sim$100~$\mu$s, this should be only a 5\% effect, but the magnetic field decay may be
slower due to eddy currents.

Finally Fig.~4 shows our observation of a maximum cloud size at
the phase transition, in agreement with the prediction of the
model.  The cloud size may not have fully equilibrated since our
ramp time was only marginally adiabatic, but this alone cannot
explain the observed maximum.

The suppression of the atom loss rate, the minimum in kinetic energy, and the maximum in cloud
size show a strong temperature dependence between $T/T_F$ of 0.12 and
0.22. As the properties of a normal Fermi gas approaching the
unitarity limit with $k_{F}^{\circ}a >> 1$ should be insensitive to
temperature variations in this range, this provides further
evidence for a transition to a new phase.

At higher temperature (e.g. $T/T_{F}=0.39$ in Fig.~3), the
observed non-monotonic behavior becomes less pronounced and shifts to larger values of $k_{F}^{\circ}
a$ for  $3\leq k_{F}^{\circ}a \leq6$. For all three observed properties (Figs.~2-4), a
nonmonotonic behavior is no longer observed at $T/T_{F}=0.55$~\cite{unitarity}. One
interpretation is that at this temperature and above, there is no
phase transition any more.  Note that in a mean-field
approximation, a ferromagnetic phase would appear at all
temperatures, but for increasing values of $k_{F}^{\circ} a$.  Our
observations may imply that the interaction energy saturates
around $k_{F}^{\circ}a \approx 5$.

The spin-polarized ferromagnetic state should not suffer from
inelastic collisions. However, typical lifetime were 10 - 20~ms, probably related to a
small domain size (see below) and three-body recombination at
domain walls.

We were unsuccessful in imaging ferromagnetic domains using
differential in-situ phase-contrast
imaging~\cite{shin:030401}.  A noise level of $S/N\sim$10 suggests
that there were at least 100 domains in a volume given by our
spatial resolution of $\sim$~3~$\mu$m and the radial size of the
cloud.  This implies that the maximum volume of the spin domains
is $\sim$~5~$\mu$m$^3$, containing $\sim$ 50 spin-polarized atoms.
We suspect that the short lifetime prevented the domains from growing to a larger size,
and eventually adopting the equilibrium texture of the ground
state, which has been predicted to have the spins pointing radially outward, like
a hedgehog~\cite{Pcoleman09,thywissen09}.
All our measurements are sensitive only to local spin polarization, independent of
domain structure and texture.

The only difference between our experiment and the ideal Stoner model is a
molecular admixture of 25 \% (Fig.~4). The molecular fraction was constant for $k_{F}^{\circ}a>$~1.8 for all temperatures
and therefore cannot be responsible for the sudden change of
behavior of the gas at $k_{F}^{\circ}a \simeq 2.2$ for the coldest temperature $T/T_{F} = 0.12$ .
This was confirmed by repeating the kinetic energy measurements with a molecular admixture of 60 \%.
The minimum in the kinetic energy occurred at the
same $k_{F}^{\circ}a$ within experimental accuracy.

Before we can compare the observed phase transition at $k_{F}^{\circ}
a\simeq 2.2$ to the theoretical predictions, we have to replace the ideal gas  $k_{F}^{\circ}$ by the value for the
interacting gas, which is smaller by $\sim$ 15\%
because of the expansion of the cloud (Fig.~4), and obtain a critical value for $k_{F}a\simeq 1.9~\pm~ 0.2$. At $T/T_{F} = 0.12$, the finite temperature
correction in the critical value for $k_{F} a$ is predicted to be less than 5\%~\cite{PhysRevLett.95.230403}.
The observed value for $k_{F} a$  is larger than the mean-field
prediction of $\pi/2$ and the second order prediction of 1.054 at zero temperature~\cite{PhysRevLett.95.230403}. Depending on the theoretical approach, the phase transition has been
predicted to be first or second order. This could not been discerned in our experiment due to the inhomogeneous
density of the cloud.

 Ref.~\cite{PhysRevLett.95.230403} speculated that earlier experiments on the measurement
 of the interaction energy~\cite{ENSexp} and RF spectroscopy of Fermi gases~\cite{S.Gupta06132003}
 showed evidence for the transition to a ferromagnetic state at or below $k_{F}a = 1$.
 This interpretation is ruled out by our experiment.

Our work demonstrates a remarkable asymmetry between positive and
negative scattering length.  Early work~\cite{PhysRevA.56.4864} predicted that
for $k_F|a|=\pi/2$, both an attractive and a repulsive Fermi
gas become mechanically unstable (against collapse, and phase
separation, respectively). In an attractive Fermi gas, however, the mechanical instability does not
occur (due to pairing~\cite{Nozieres85}), in contrast to our observations in a repulsive Fermi gas.
This suggests that the maximum total repulsive energy (in units of
$3/5{(2Vn)}E_F$) is larger than the maximum attractive energy $|\beta|$ of
0.59~\cite{PhysRevLett.95.060401}
realized for infinite $a$~\cite{exp}.

Heisenberg's explanation for ferromagnetism was based on exchange
energy, i.e. the Pauli principle and spin-independent repulsive interactions
between the electrons. However, it remained an open question, what other
``ingredients" were needed for itinerant ferromagnetism.  It was only in 1995~\cite{PhysRevLett.75.4678,tanaka:116402},
 that a rigorous proof was given that, in certain lattices, spin-independent Coulomb interactions can give rise to ferromagnetism
  in itinerant electron systems. Our finding implies that Heisenberg's idea does not require a lattice and band
 structure, but applies already to a free gas with short-range interactions. Our experiment can be regarded as quantum simulation of a Hamiltonian
for which even the existence of a phase transition was unproven.  This underlines
the potential of cold atom experiments as quantum simulators for many-body physics.


\clearpage

\begin{figure}
\begin{center}
\includegraphics[width=4.5in]{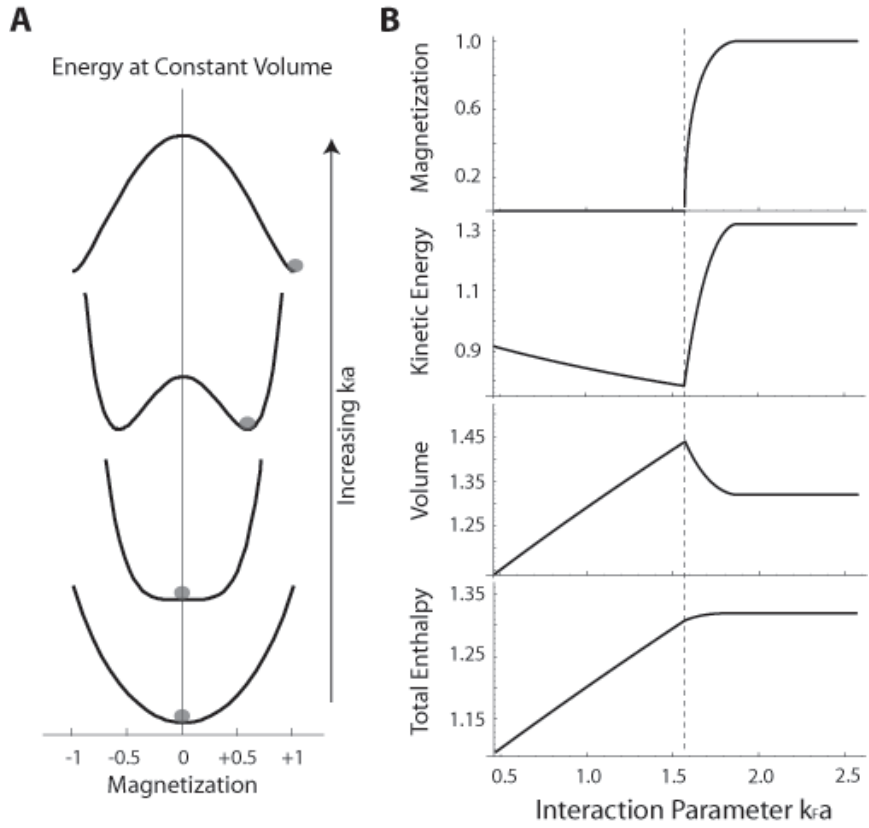}
\caption{Ferromagnetic phase transition at T=0, according to the
mean-field model described in the text.  The onset of itinerant
ferromagnetism occurs when the energy as a function of
magnetization flips from a U-shape to a W-shape (\textbf{A}).  Figure (\textbf{B})
shows the enthalpy, volume and kinetic energy (normalized to their
values for the ideal Fermi gas), and magnetization as a function
of the interaction parameter $k_F a$.  Note that $k_F$ is defined
by the density of the gas. The dotted line marks
the phase transition.}
\end{center}
\end{figure}

\clearpage
\begin{figure}
\begin{center}
\includegraphics[width=4.5in]{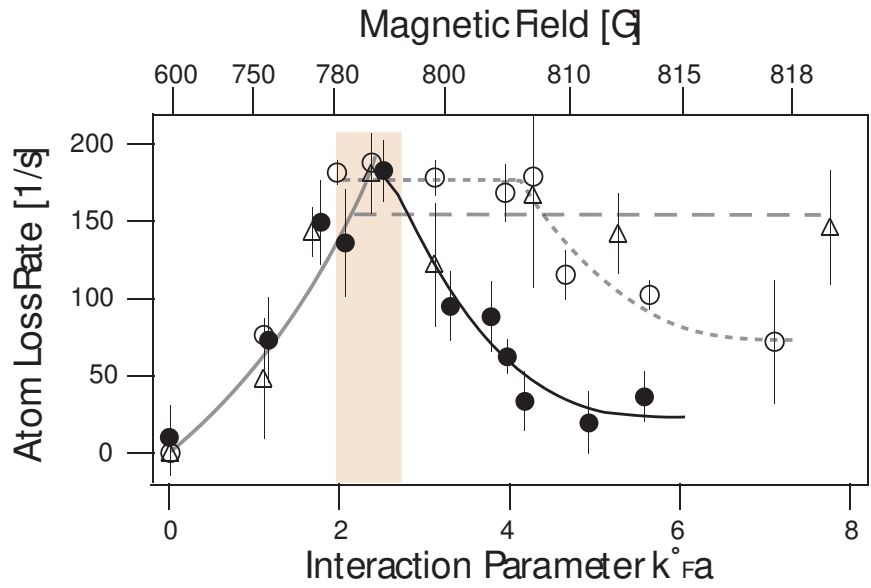}
\caption{Atom loss rate as a probe for local spin
polarization, for different temperatures.  (a) $T/T_F=0.55$
(dashed curve), (b) $T/T_F=0.22$ (dotted curve), and $T/T_F=0.12$
(solid black curve). The curves are guides to the eye, based on the assumption of a loss rate
 which saturates for increasing $a$ in the normal state. The shaded area around
the phase transition at $T/T_F=0.12$ highlights the same region as in Figs. 3 and 4.}
\end{center}
\end{figure}

\clearpage

\begin{figure}
\begin{center}
\includegraphics{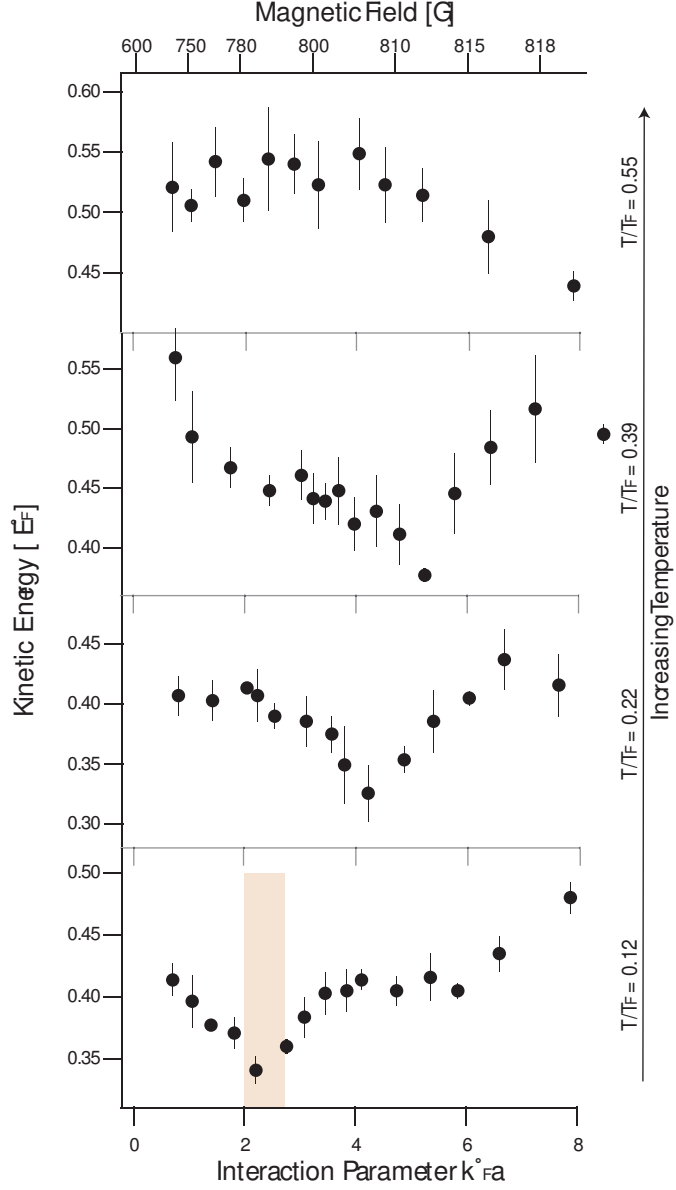}
\caption{Kinetic energy of a repulsively
interacting Fermi gas determined for different interaction
parameters $k_{F}^{\circ}a$ and temperatures. The measured kinetic energy
is normalized by the Fermi energy $E_{F}^{\circ}$ of the noninteracting
Fermi gas at $T$=0, calculated at the trap center with the same number of atoms per spin state.
 Each data point represents the average of
three or four measurements. }
\end{center}
\end{figure}

\clearpage

\begin{figure}
\begin{center}
\includegraphics{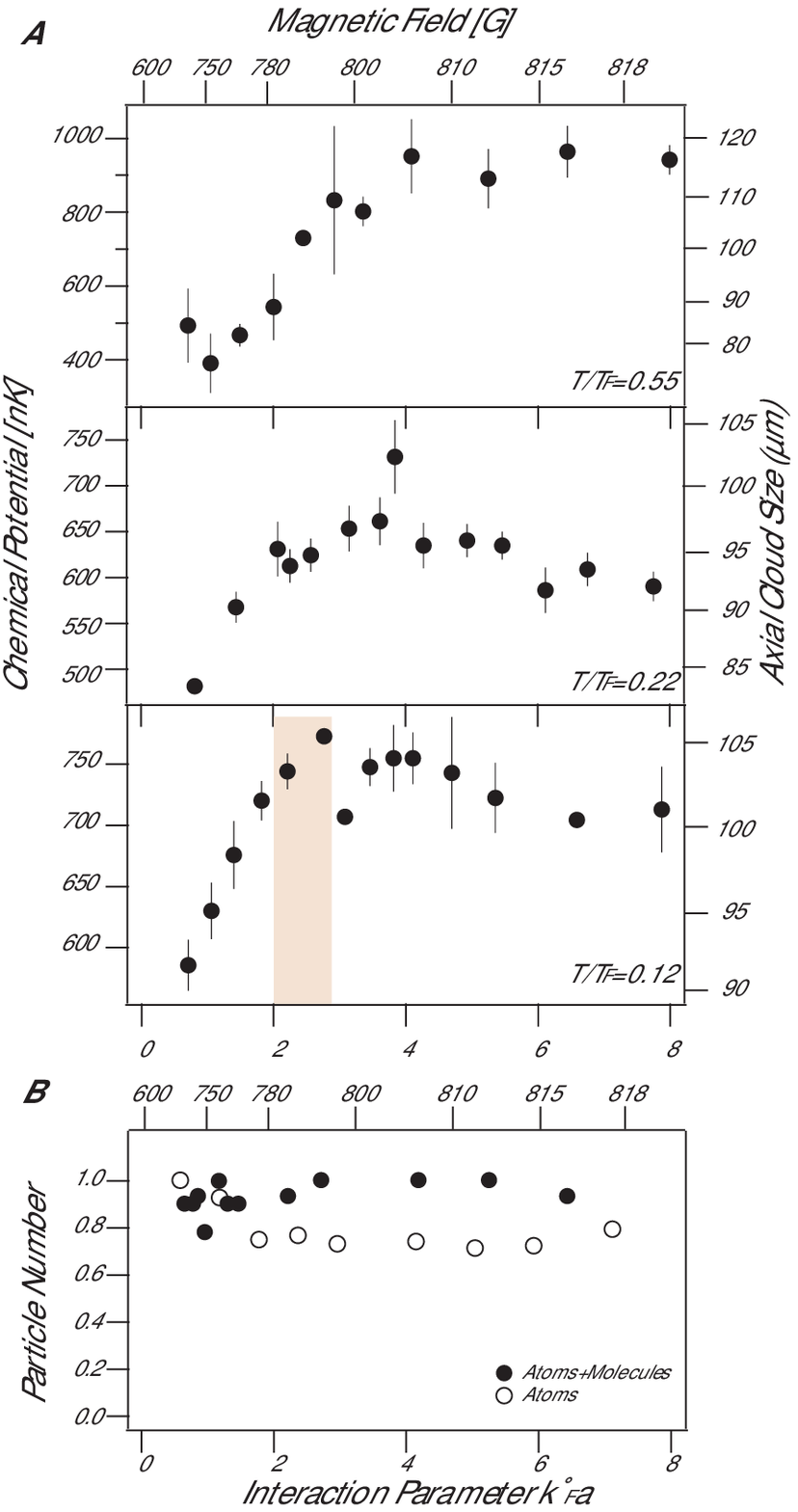}
\caption{Maximum in volume at the phase transition
(\textbf{A}) Axial size and chemical potential of the cloud for various
temperatures.  The chemical potential $\mu$ is determined from the
measured cloud size, $\sigma_{z}$ as
$\mu=\frac{1}{2}m{\omega_{z}}^2{\sigma_{z}}^2$ (\textbf{B}) Number of
particles including both atoms and molecules right after the
field ramp.  This result
shows that 25 \% of atoms were converted into molecules during the
field ramp, and this fraction stayed constant for $k_{F}^{\circ}a>1.8$,
where the phase transition was reached.  This molecule fraction
was independent of temperature.}
\end{center}
\end{figure}

\clearpage


\begin{thebibliography}{10}
\bibitem{unitarity}
The interpretation of the loss rate is complicated since $f(a,T)$ is unkown for
  $k_{F}a \geq 1$. The three body rate $f(a,T)$ is expected to be unitarity
  saturated for $k_{F}a \gg 1$~\cite{PhysRevLett.91.123201}. The lines in Fig.
  2 indicate that the observed loss rate is consistent with unitarity
  saturation and a sudden drop at the phase transition, which occurs at large
  values of $k_{F}a$ at higher temperature.
\end{thebibliography}

\bibliographystyle{Science}

\begin{scilastnote}
\item
This work was supported by the NSF and ONR, through a MURI
program, and under ARO Grant No. W911NF-07-1-0493 with funds from
the DARPA OLE program.  G.-B. Jo and Y.-R. Lee acknowledge
additional support from the Samsung Foundation. We would like to
thank  E. Demler, W. Hofstetter, A. Paramekanti, L. J. LeBlanc, and
G.J. Conduit for useful discussions, T. Wang for experimental assistance,
and  A. Keshet for development of the computer control system.
\end{scilastnote}

\newpage


\section*{Supporting materials: \\
Itinerant Ferromagnetism in a Fermi Gas of Ultracold Atoms\\
}

\subsection*{Materials and Methods}
\paragraph*{Preparation of the ultracold $^6$Li cloud}
The first step is the production of a spin-polarized  Fermi
gas in the $|F=3/2,m_{F}=3/2\rangle$ state by sympathetic cooling
with bosonic $^{23}$Na atoms in a magnetic trap as described
in ref~({\it S1}). The $^6$Li cloud was then loaded
into a deep optical dipole trap with a maxium power of 3W and radial
trap frequency of $\sim$3.0~kHz, followed by an RF
transfer into the lowest hyperfine state $|F=1/2,m_{F}=1/2\rangle$. Additional axial confinement was provided by
magnetic fields. An equal mixture of $|1\rangle$
and $|2\rangle$ spin states (corresponding to the $|F=1/2,m_{F}=1/2\rangle$ and $|F=1/2,m_{F}=-1/2\rangle$ states at low magnetic
field) was prepared by a Landau-Zener RF
sweep at a magnetic field of 590 G, followed by 1~s for
decoherence and further evaporative cooling at ~300 G. Finally,
the optical trapping potential was adiabatically reduced over
600~ms, and the field increased back to 590 G. The trap had a depth of 7.1 $\mu$K and was
nearly cigar shaped with frequencies
$\nu_{x}=\nu_{y}\simeq 300$~Hz and
$\nu_{z}\simeq70$~Hz.


\subsection*{Supporting online text}
\paragraph*{Estimation of the maximum total repulsive energy}
Full phase separation at zero
temperature requires a total repulsive energy of $(2^{2/3}-1)=0.59$ in units of
$3/5{(2Vn)}E_F$. At finite temperature $T$, one has to add $ T S$ where $S=(2Vn)k_B ln2$ is the
entropy difference between the two phases. Our
tentative observation of a ferromagnetic phase at $T=0.39 T_F$
implies a repulsive energy of $\sim 1.04$ assuming full phase separation, larger than the maximum attraction
energy of 0.59.

\clearpage

\begin{figure}
\begin{center}
\includegraphics[width=4.0in]{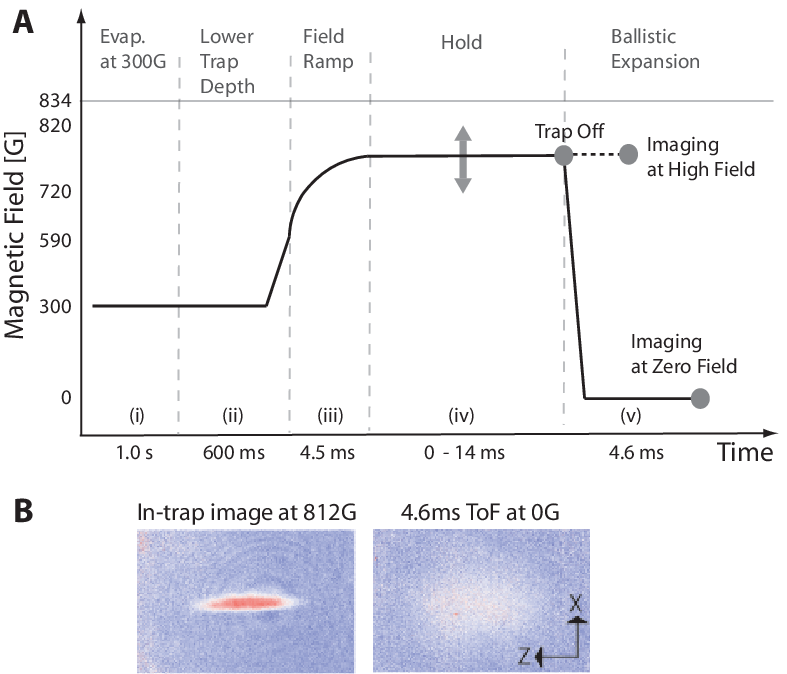}
\end{center}
\end{figure}
\textbf{Fig. S1}: (\textbf{A}) The schematic shows the time sequence
of the experiment. The sample was exposed to the
magnetic field of interest for 0 - 14 ms and analyzed in-situ for loss measurement or
after 4.6~ms time-of-flight for the measurement of kinetic energy and the axial size of the cloud.
The Feshbach fields were suddenly
switched off at a rate of 1G$/{\mu}$s, preventing the conversion
of interaction energy into kinetic energy during the expansion.
(\textbf{B}) This absorption image shows the $|1\rangle$ component of
the cloud  trapped at 812~G (left), and after 4.6~ms ballistic
expansion imaged at zero field (right). The field of view is
$840\mu$m$\times550\mu$m. The magnetic field ramp was limited by eddy currents to 4.5~ms.
Spectroscopic measurements of the magnetic field showed that the field was trailing behind the current
which was controlled with a time constant faster than 1~ms.

\subsection*{References and Notes}
\begin{itemize}
\item[S1.]
Z.~Hadzibabic, {\it et~al.\/}, {\it Phys. Rev. Lett.\/} {\bf 91}, 160401
  (2003).

\end{itemize}

\end{document}